\documentclass[12pt,preprint]{aastex}
\newcommand{\myemail}{maggie@physics.mcgill.ca}
\newcommand{\psr}{PSR~J1846$-$0258}

\newcommand{\rxte}{{\textit{RXTE}}}

\newcommand{\nudotdotdot}{{\ifmmode\stackrel{\bf \,...}{\textstyle \nu}\else$\stackrel{\,\...}{textstyle \nu}$\fi}}
\newcommand{\degrees}{^{\circ}}
\newcommand{\asca}{{\it ASCA}}
\newcommand{\beppo}{{\it BeppoSAX}}

\usepackage{graphicx}

\shorttitle{A Braking Index for \psr}
\shortauthors{Livingstone et al.}

\begin{document}
\title{A Braking Index for the Young, High-Magnetic-Field, Rotation-Powered Pulsar in
Kes 75}

\author{Margaret A.~Livingstone \altaffilmark{1},
Victoria M.~Kaspi}
\affil{Department of Physics, Rutherford Physics Building, 
McGill University, 3600 University Street, Montreal, Quebec,
H3A 2T8, Canada}

\author{E.~V.~Gotthelf}
\affil{Columbia Astrophysics Laboratory, Columbia University, 550 West 120th
Street, New York, NY 10027-6601}
\and
\author{L.~Kuiper}
\affil{Netherlands Institute for Space Research, Sorbonnelaan 2, 3584 CA,
Utrecht, Netherlands}
\altaffiltext{1}{\myemail}

\clearpage

\begin{abstract}
We present the first phase-coherent measurement of a braking index
for the young, energetic rotation-powered pulsar \psr. This 324\,ms pulsar
is located at the center of the supernova remnant Kes 75 and has a
characteristic age of $\tau_c = 723$\,years, a spin-down energy of
$\dot{E}=8.3 \times 10^{36}$\,erg\,s$^{-1}$, and inferred magnetic field of
$4.9\times 10^{13}$\,G. Two independent
phase-coherent timing solutions are derived which together span 5.5\,yr
of data obtained with the {\textit{Rossi X-ray Timing Explorer}}. 
In addition, a partially phase-coherent timing analysis confirms
the fully phase-coherent result. The measured value of the braking 
index, $n=2.65\pm0.01$, is significantly less than 3, the value 
expected from magnetic dipole radiation, implying another physical process
must contribute to the pulsar's rotational evolution. 
Assuming the braking index has been constant since birth, 
we place an upper limit on the spin-down age
of \psr\ of 884\,yr, the smallest age estimate of any rotation-powered
pulsar. 
\end{abstract}

\keywords{pulsars: general---pulsars: individual (\objectname{\psr})---X-rays: stars}

\section{Introduction}
\label{sec:intro}
The measurement of pulsar braking indices ($n$) is crucial to the understanding of
the physics underlying pulsar spin down, assumed to be of the form
\begin{equation}
\dot{\nu} = -K \nu^n,
\end{equation}
where $\nu$ is the pulse frequency, 
$\dot{\nu}$ is the frequency derivative and $K$, assumed to be constant, is related to the
pulsar's magnetic field and moment of inertia. By taking a time derivative of the above 
spin-down equation, we find that 
$n = {\nu \ddot{\nu}}/{\dot{\nu}^2}$, where $\ddot{\nu}$ is the second
frequency derivative. It is typically assumed that pulsars radiate as
perfect magnetic dipoles, implying $n=3$. This assumption is used implicitly for the
estimation of pulsar magnetic fields as well as to calculate characteristic
ages (defined as $\tau_c=P/2{\dot{P}}$ corresponding to $n=3$). However, 
of the five unambiguous measurements of pulsar braking indices
obtained so far, all yield a value of $n<3$
\citep{lps93,lpgc96,ckl+00,lkg05,lkgm05}.
Explanations for this discrepancy
include the possibility that: the relativistic pulsar wind affects the 
spin-down \citep{mt69};
the pulsar may suffer a propeller torque from a putative supernova
fallback disk \citep{mph01}; the pulsar has
a time-varying magnetic moment \citep{br88}; or, the pulsar cannot
be modelled as a point dipole, rather, a dipole of some finite size which
leads to $n<3$ \citep{mel97}. 

There are few pulsars that are potential candidates for a
significant measurement of $n$; 4 of the 5 measured values are from sources
with characteristic ages under 2000\,yr. The first requirement for a
significant measurement of $n$ is that the pulsar spins down sufficiently
quickly for a measurement of $\ddot{\nu}$. The second requirement is that
the position of the pulsar is accurately known at the $\sim$1 arcsecond 
level. The third requirement is that
the spin-down must not be seriously affected by glitches,  sudden spin-ups 
of the pulsar, or timing noise, a long-term,
low-frequency stochastic wandering of the rotation about the overall trend. 
Typically, glitches begin
to seriously affect smooth spin down at characteristic ages of $\sim 5-10$\,kyr
\citep{ml90,mgm+04}. Thus many of the pulsars that may spin down fast enough for a 
measurement of $n$ are irretrievably contaminated by glitches
\citep[e.g.][]{mgm+04}. Timing noise
varies from object to object, though is roughly correlated with spin-down
rate \citep[e.g.][]{antt94} and can prevent a measurement of
$n$ in a finite data set in an unpredictable way.

The very young, energetic pulsar \psr\ was discovered at the center of the
supernova remnant Kes 75 \citep{gvb+00}. Neutral hydrogen absorption
measurements indicate that the pulsar resides well across the Galaxy,
roughly at a distance of 19\,kpc \citep{bh84}. 
Radiation from this pulsar is detected exclusively in the
X-ray band, with a 0.5-10\,keV luminosity of $L_x=4.1 \times 10^{35}$\,erg\,s$^{-1}$
for a distance of 19 kpc \citep{hcg03}. If correct, the spin-down energy to 
X-ray conversion efficiency is 1.6\%, 6
times greater than that found in the Crab pulsar. The X-ray spectrum is
typical of rotation-powered pulsars, with non-thermal power-law emission
with $\Gamma =1.4$ \citep{hcg03}. However, its large inferred magnetic
field of $B = 4.9 \times 10^{13}$\,G and spectrum place it in
an emerging class of rotation-powered pulsars with magnetar-strength fields
\citep[e.g.][]{km05}. With a characteristic age of $\tau_c = 723$\,yr, \psr\
is likely the youngest of all known rotation-powered pulsars. 

In this paper, we present the first phase-coherent timing solution for the
Kes 75 pulsar based on 5.5\,yr of X-ray timing data from a long-term
monitoring campaign with the {\textit{Rossi X-ray Timing Explorer}} (\rxte).

\section{\rxte\ Observations and Analysis}
\label{sec:obs}
Observations of \psr\ were made using the Proportional Counter Array 
\citep[PCA;][]{jsg+96} on board {\textit{RXTE}}. The PCA consists of an
array of five collimated xenon/methane multi-anode proportional counter
units (PCUs) operating in the 2~--~60~keV range, with a total effective
area of approximately $\rm{6500~cm^2}$ and a field of view of 
$\rm{\sim 1^o}$~FWHM.

\begin{figure}
\plotone{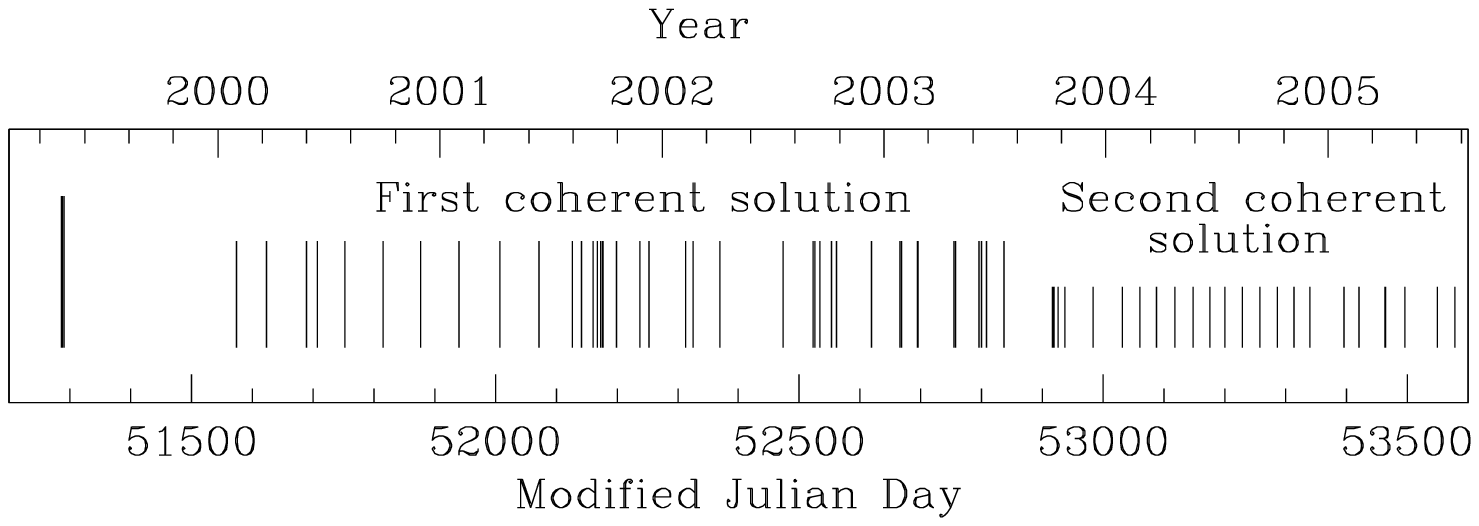}
\figcaption[f1.eps]{Distribution of \rxte\ observations of \psr\ 
over the 6.3\,yr interval. Each 
bar represents all observations occurring on a single day. 
81 merged observations are shown in 
total. The mid-length bars represent data included in the first 
phase-connected solution, while the short bars represent data included in the
second phase-connected solution. The initial \rxte\ observations (tallest
bars) of the source occur significantly before the monitoring observations 
and cannot be unambiguously phase connected with the later data. 
\label{fig:distribution}}
\end{figure}

The data span 6.3\,yr from 1999 April 18 - 2005 July 27 and were collected 
in ``GoodXenon'' mode, which records the arrival time
(with 1-$\mu$s resolution) and energy (256 channel resolution) of every
unrejected event. Typically, 3 PCUs were operational
during any observation. We used all layers of each operational PCU
in the 2--60~keV range, as this maximizes the signal-to-noise ratio
for this source. We
analyzed 171 observations, resulting in 158 detections of the
pulse for a total integration time of 337\,hr. The data were unevenly spaced
throughout the 6.3\,yr of observations, as shown in Figure~\ref{fig:distribution}. 
Because \psr\ was occasionally not the primary target of \rxte\ in the
observations we have used, integration times ranged from $\sim$1.5 to $>$25\,ks,
resulting in a variety of signal-to-noise ratios for individual pulse
profiles. 

\begin{figure}
\plotone{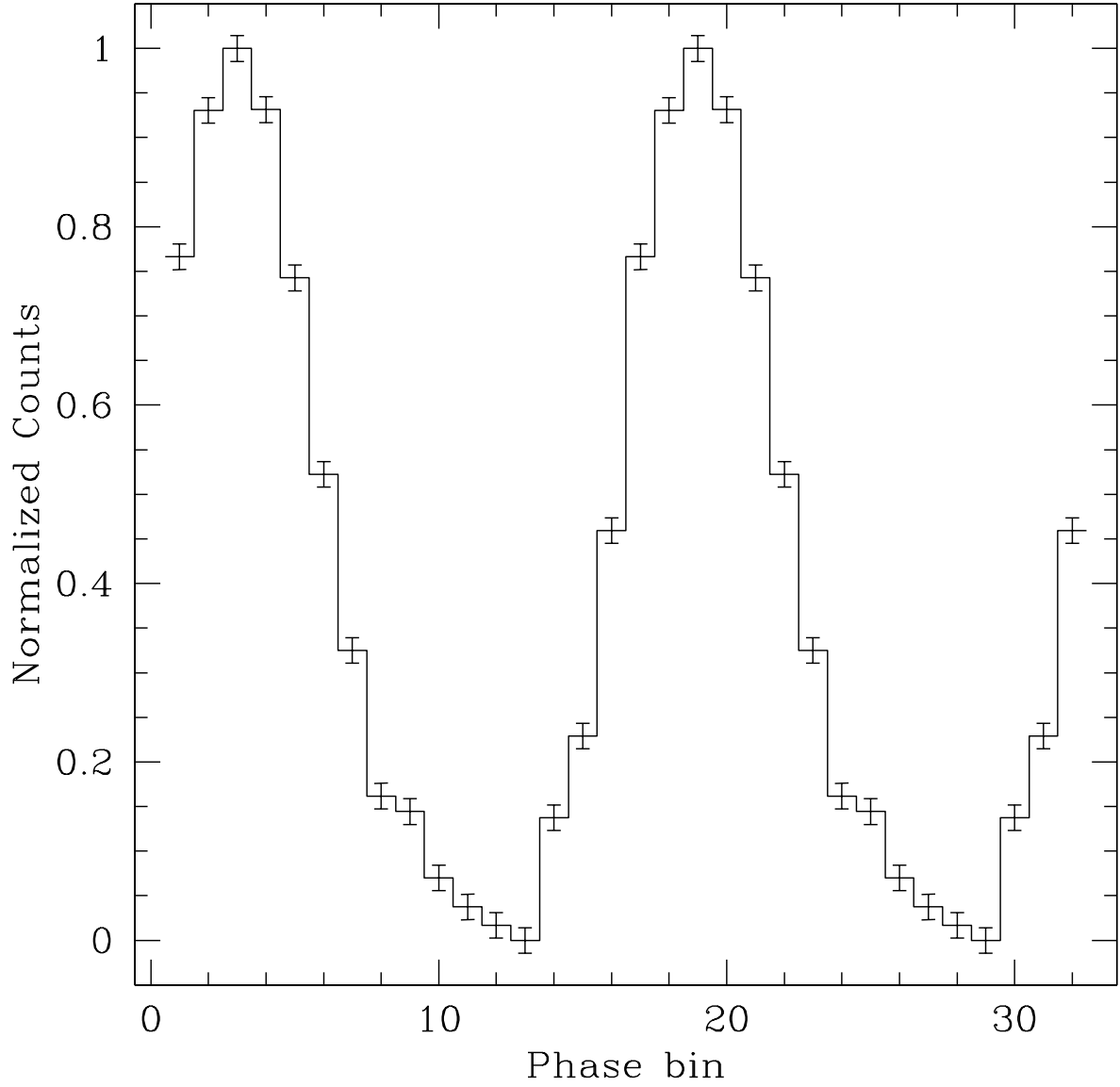}
\figcaption[f2.eps]{Integrated pulse profile of 337\,hr of data over
6.3-yr of \rxte\ observations of \psr. Two full pulse cycles are shown for clarity. 
\label{profile}}
\end{figure}

Observations were downloaded from the HEASARC 
archive\footnote{http://heasarc.gsfc.nasa.gov/docs/archive.html}
and data from the different PCUs were merged and binned at
(1/1024)\,ms resolution using the
Ftools\footnote{http://heasarc.gsfc.nasa.gov/docs/software/ftools/} `seextrct' and 
`fselect'. Photon arrival times were converted
to barycentric dynamical time (TDB) at the solar system barycenter using the
J2000 source position RA = $18^{\rm{h}}46^{\rm{m}}24\fs94$, Dec $= -02 \degrees
58\arcmin30.1\arcsec$ \citep{hcg03} and the JPL DE200 solar system ephemeris
with the FITS tool `faxbary'. The remainder of the analysis was performed 
using software developed at MIT for handling \rxte\ data.
We determined an initial ephemeris by merging  
adjacent short observations and performing a periodogram analysis that
resulted in a set of 134 frequency measurements spanning the 6.3-yr monitoring
interval. From these measurements, we derived an initial frequency
derivative, $\dot{\nu}$. This ephemeris was used to fold
each time series with 16 phase bins. This number of bins was chosen because
of the lack of features in the roughly sinusoidal profile and the resulting
reasonable 
signal-to-noise ratio for individual profiles, particularly those made from
short integrations. Resulting profiles were cross-correlated 
in the Fourier domain with a high signal-to-noise ratio template created by 
adding phase-aligned profiles from all observations, 
shown in Figure~\ref{profile}. The cross-correlation process assumes that the
pulse profile is stable; indeed, we found no evidence for variability. 
We implemented
a Fourier domain filter by using only the first six harmonics in the
cross-correlation. For each observation, the cross-correlation yeilded
the time of arrival (TOA) of phase-zero of the average pulse profile at
the fold epoch. The TOAs were fitted to a timing model (see \S \ref{sec:coherent} 
and \S \ref{sec:partial}) using the pulsar timing software package
TEMPO\footnote{http://www.atnf.csiro.au/research/pulsar/tempo}.
After phase-connecting the data (see \S \ref{sec:coherent}), we
merged observations occurring on a single day and 
used the ephemeris to re-fold the data in order to obtain more
accurate TOAs. This process produced 81 TOAs with a typical uncertainty $\sim$8\,ms
($\sim$2.5\% of the pulse period), improved from a typical TOA uncertainty
of $\sim$10\,ms. 

\subsection{Phase-Coherent Timing Analysis}
\label{sec:coherent}
   
To obtain a phase-coherent timing solution, each turn of the 
pulsar is accounted for by fitting TOAs to a Taylor expansion of pulse
phase, $\phi$. At time $t$, $\phi$ can be expressed as: 
\begin{equation}
\phi(t)=\phi(t_0) + \nu_0(t-t_0) + \frac{1}{2}{\dot{\nu}_0}(t-t_0)^2 +
        \frac{1}{6}{\ddot{\nu}_0}(t-t_0)^3 + {\ldots}  ,
\end{equation}
where the subscript 0 denotes a parameter evaluated at the reference
epoch, $t_0$. TOAs and initial parameters are input into TEMPO, 
which gives as output refined spin parameters and residuals. 

To determine the spin parameters for \psr, we obtained phase-connected timing
solutions for 5.5\,yr of X-ray timing data spanning 2000 January 31 to
2005 July 27, and including 78 TOAs.
The sampling of the observations over 6.3\,yr includes several large gaps and
that precludes a single phase-coherent timing solution for the entire interval.

The first \rxte\  observations of the pulsar (which resulted in its discovery)
occurred $\sim$9 months prior to the commencement of regular monitoring
observations, and so were not useful in our phase-coherent analysis. The 
first coherent solution is valid over MJD 51574-52837
(3.5\,yr), while the second coherent solution is valid over MJD 52915-53578
(1.8\,yr), as indicated in Figure~\ref{fig:distribution}.

\begin{figure}
\plotone{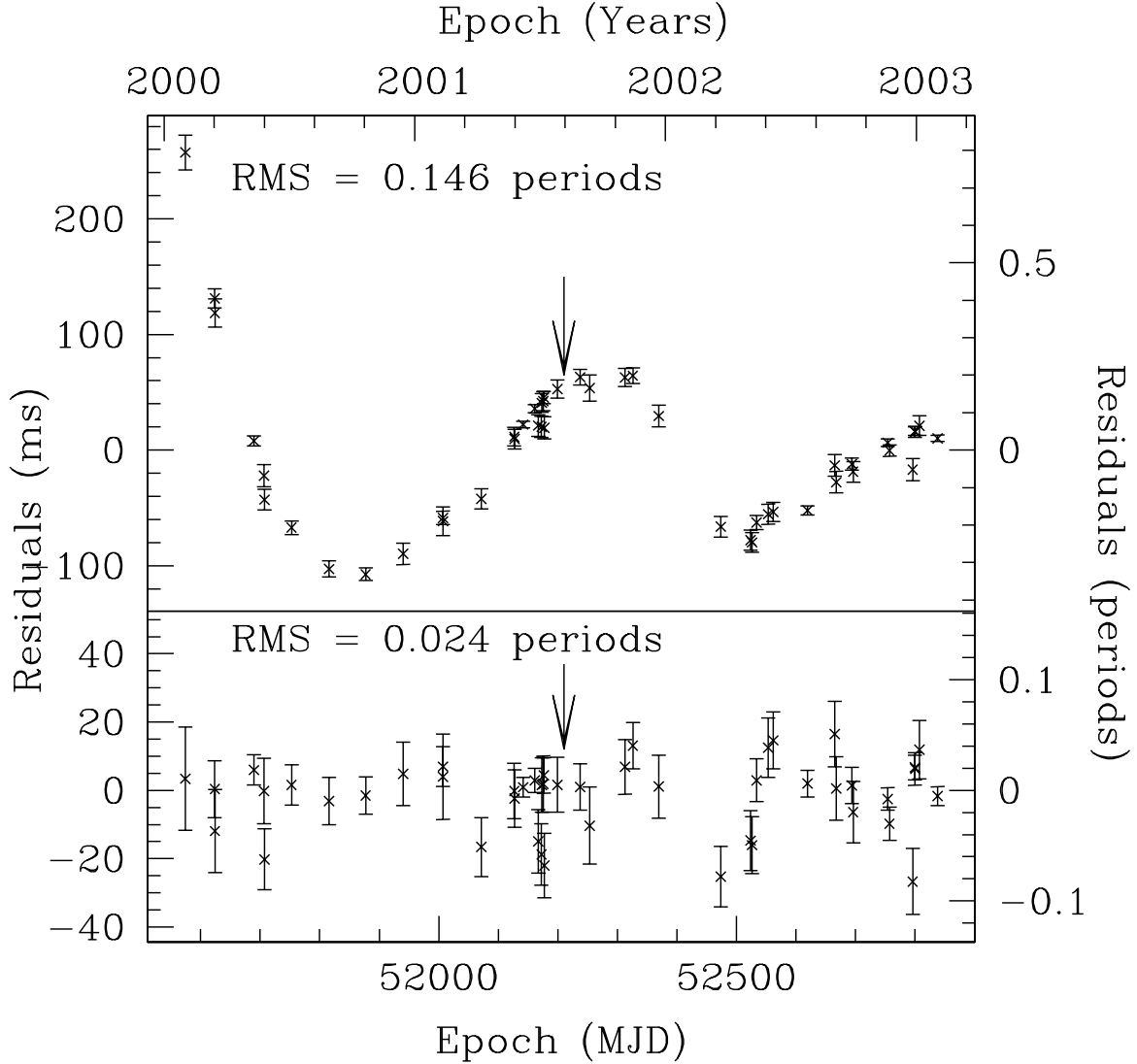}
\figcaption[f3.eps]{Phase-coherent X-ray timing analysis of the
young pulsar \psr\ spanning a 3.5-yr interval from MJD 51574
-52837. Top panel:
Residuals with $\nu$, $\dot{\nu}$, $\ddot{\nu}$, as well as glitch parameters
$\Delta{\nu}$ and $\Delta{\dot{\nu}}$ fitted. The glitch epoch, MJD 52210 is
indicated by the arrow. Bottom panel:
Residuals with glitch parameters and eight frequency derivatives in total fitted
to render the residuals consistent with Gaussian noise. Fitting the additional
parameters improves the $\chi^2$ from 2933 for 43 degrees of freedom to 77 for
37 degrees of freedom. Although these parameters do not completely describe the data, 
the $\chi^2$ is not improved significantly by fitting additional frequency
derivatives, which is not uncommon when timing noise is present.
\label{fig:ephem1}}
\end{figure}

We used our initial ephemeris (described in the previous section) to
bootstrap a phase-coherent solution, valid over the 3.5\,yr interval from 
MJD 51574-52837. This solution includes $\nu$, $\dot{\nu}$ and 
$\ddot{\nu}$, whose values are given in Table 1. In the process of phase
connection, it became clear that a small glitch occurred
at MJD 52210$\pm$10. Our measured glitch parameters are $\Delta{\nu}/\nu = 2.5(2) \times
10^{-9}$ and $\Delta{\dot{\nu}}/{\dot{\nu}} \sim 9.3(1)\times 10^{-4}$,
as determined with the glitch fitting facility in TEMPO. The relatively wide
spacing of data near the glitch epoch prevent the detection of any glitch
recovery. In fact, it is possible that the initial frequency jump was
larger and recovered significantly before the following observation. 
Timing residuals after subtraction of our best-fit timing model, including
the glitch, are shown in the top panel of Figure~\ref{fig:ephem1}.
Note that systematic trends remaining in the residuals are likely the result of
timing noise, common among young pulsars, though unmodelled glitch recovery
may also contribute to the observed residuals. Timing noise processes are 
known to contaminate measured spin
parameters, hence it is typically advisable to remove the systematics from 
the residuals by fitting additional frequency derivatives until the residuals
are consistent with Gaussian noise \cite[e.g.][]{kms+94}. 
For this pulsar, a total of eight frequency derivatives were required to
obtain Gaussian distributed residuals. Timing residuals with all eight
derivatives removed are shown in the bottom panel of
Figure~\ref{fig:ephem1}. Fitting additional frequency derivatives 
improves the $\chi^2$ from
2933 for 43 degrees of freedom to 77 for 37 degrees of freedom.
This $\chi^2$ value indicates that the fit does not completely
describe the data, however, this is not uncommon when fitting 
timing noise, which is often not well described by a polynomial. 
The braking index, resulting from this `whitened' 
timing solution, is $n=2.64\pm 0.01$. 
Deterministic spin-down parameters (i.e. not the higher order derivatives
that represent timing noise) as well as glitch parameters 
for this timing solution are given in Table~\ref{table:ephem}. 

Phase was lost over a 78-day gap in the data beginning at MJD 52837, made
clear by the fact that 
a solution attempting to connect over this gap fails to predict the pulse
frequency at previous epochs. 
This loss of phase could be due either to timing noise or another glitch. 
However, the estimated change in
frequency over the gap, calculated from phase-coherent solutions on each
side of the gap, is negative, ie. in the opposite direction from 
a conventional glitch. This implies that a glitch 
alone cannot account for the loss in phase. 

\begin{figure}
\plotone{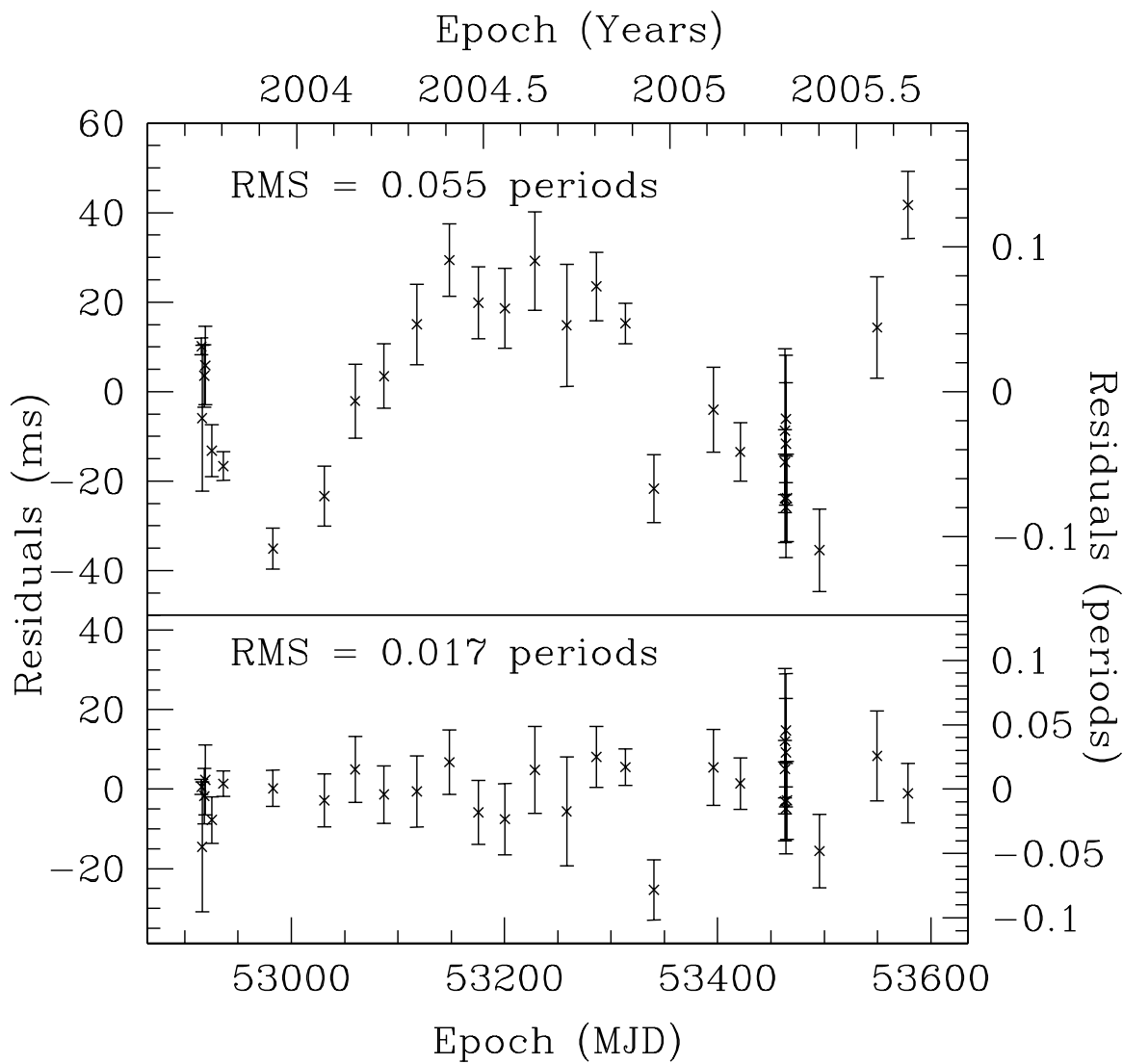}
\figcaption[f4.eps]{Phase-coherent X-ray timing analysis of
\psr\ spanning a 1.8-yr interval from MJD 52915-53579. Top
panel: Residuals with $\nu$, $\dot{\nu}$, and $\ddot{\nu}$ fitted. Bottom
panel: Residuals with five frequency derivatives total fitted to render the residuals
consistent with Gaussian noise. Fitting the additional parameters improves
the $\chi^2$ from 271 for 27 degrees of freedom to 26 for 24 degrees of
freedom.\label{fig:ephem2}}
\end{figure}

A second phase-coherent solution was obtained for the 1.8-yr interval from
MJD 52915-53579, with
$\nu$, $\dot{\nu}$, $\ddot{\nu}$ fitted. The residuals after subtraction of
the best-fit parameters are shown in the top panel of
Figure~\ref{fig:ephem2}.
As systematic trends, again interpreted as timing noise or possibly
unmodelled glitch recovery, remained in the
residuals, three higher-order frequency derivatives were fitted. The
resulting residuals are shown in the bottom panel 
of Figure~\ref{fig:ephem2}. The measured braking index resulting from the `whitened' timing 
solution is $n=2.68\pm 0.03$, in agreement with that measured
from the first segment of timing data. Deterministic parameters for the second timing
solution are given in Table~\ref{table:ephem}. 

\subsection{Partially Phase-Coherent Timing Analysis}
\label{sec:partial}
In order to mitigate the effects of timing noise,
we also performed a partially phase coherent analysis. In this way
we obtained measurements of $\nu$ spanning 6.3\,yr of data as well as $\dot{\nu}$ and
$\ddot{\nu}$ over 5.5\,yr of data. This method is useful to detect
small glitches, as well as to obtain more accurate measurements of
$n$ in some cases \citep[e.g.][]{lkg05}. Using the overall phase-coherent ephemeris as a
starting point, closely spaced observations were phase-connected
to obtain a local measurement of $\nu$. In this way we obtained 
a total of 22 measurements. 
A two degree polynomial was fitted to these measurements
to get another measurement of $n$, however, in this case, the glitch
near MJD 52210 and the possible glitch between MJD 52837-52915 seriously
restricted the available time baseline and rendered this analysis of limited
value. The most constraining measurement is $n=2.83\pm0.39$, from 13 values 
of $\nu$ spanning MJD 51286-52199, in agreement with our phase-coherent value. 

\begin{figure}
\plotone{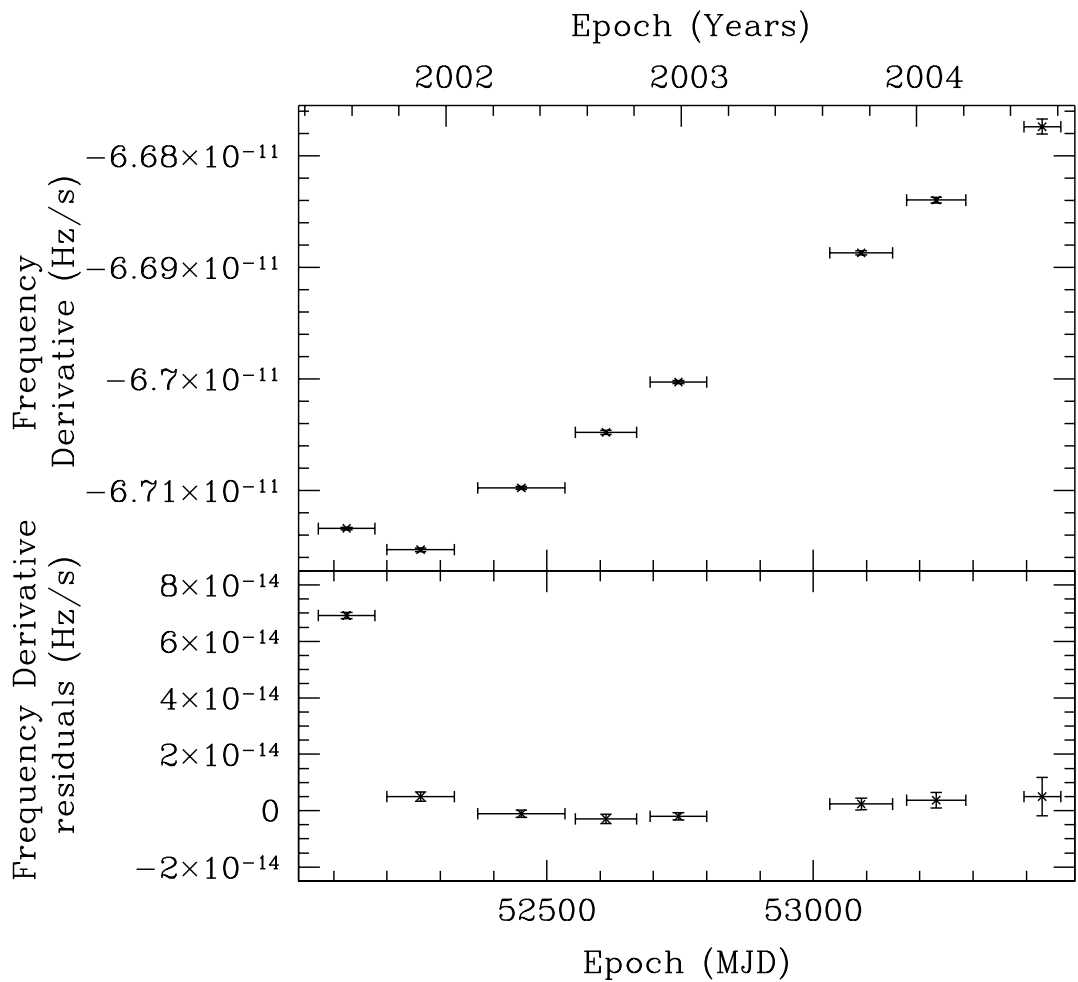}
\figcaption[f5.eps]{Eight phase-coherent $\dot{\nu}$ measurements
spanning 5.5\,yr of \rxte\ monitoring observations of \psr. Top panel:
Measurements of $\dot{\nu}$. There is a clear discontinuity
after the first measurement of $\dot{\nu}$, which we interpret as a glitch.
Bottom panel: $\dot{\nu}$ measurements with the post-glitch slope fitted to
highlight the discontinuity, which is quantified as
$\Delta{\dot{\nu}}/{\dot{\nu}}= (9.5\pm 0.3) \times 10^{-4}$.  
\label{fig:nudot}}
\end{figure}

We repeated this process, measuring eight independent values of $\dot{\nu}$, 
shown in the top panel of Figure~\ref{fig:nudot}. 
Note that the first unambiguous $\dot{\nu}$ measurement occurred 
before the glitch near MJD 52210 and the glitch can clearly be seen in the
Figure. The bottom panel of Figure~\ref{fig:nudot} shows the post-glitch slope removed from the 
data, highlighting the change in $\dot{\nu}$ at the time of the glitch, 
$\Delta{\dot{\nu}}/{\dot{\nu}}= (9.5\pm 0.3) \times 10^{-4}$, 
in agreement with the value obtained from the phase-coherent fit.

Note that at the 78\,day data gap between MJD 52837 and 52915,
there is no clear $\Delta{\dot{\nu}}$ visible in 
Figure~\ref{fig:nudot}. This indicates that if a glitch did occur during this
period, it consisted of primarily a change in $\nu$.
The only glitch consistent
with the data would have had to have occurred near MJD 52910 with magnitude 
$\Delta{\nu}/{\nu} < 5 \times 10^{-8}$. A glitch with no change in
$\Delta{\dot{\nu}}$ is typical of pulsars with larger characteristic ages 
($\ga 5$\,kyr), but cannot be ruled out in this case. 

We also performed two linear weighted least square fits
on the $\dot{\nu}$ measurements, both before and after the gap
at MJD 52837-52915, to obtain measurements of $\ddot{\nu}$ and thus
$n$. The first segment resulted in $n=2.47\pm 0.07$, while the second
resulted in $n=2.65\pm 0.14$, $1.3\sigma$ from each other. Uncertainties
were obtained from a bootstrap analysis, since we suspected that the formal
uncertainty may underestimate the true uncertainty because of the presence
of timing noise. The bootstrap is a robust method of determining errors
when a small number of sample points is available, as in this case
\citep{efr79}.
The first measurement is $2.5\sigma$ from our phase-coherent average value
of $n=2.65\pm0.01$, while the second measurement is in agreement with it.

\begin{figure}
\plotone{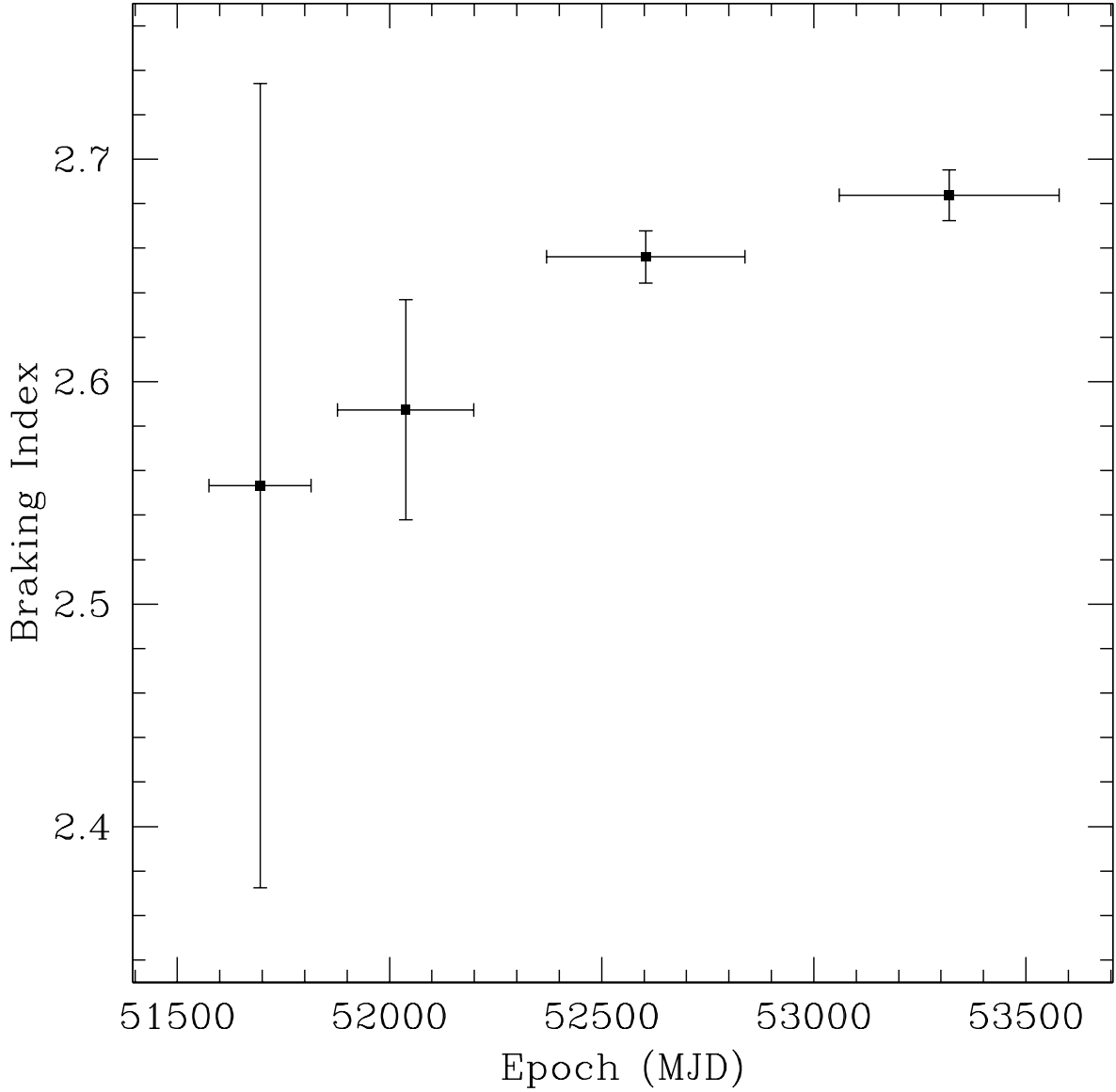}
\figcaption[f6.eps]{Phase-coherent measurements of the braking index,
$n$ of \psr. Each phase-connected solution has only $\nu$, $\dot{\nu}$ and
$\ddot{\nu}$ fitted and has Gaussian residuals. The braking index is
consistent with being constant, since the slight increasing trend of
$\Delta{n}/\Delta{t} \simeq 0.02\pm0.01$\,yr$^{-1}$ is only
significant of the 2$\sigma$ level. 
\label{fig:indices}}
\end{figure}

Finally, we obtained four independent measurements of $\ddot{\nu}$
by phase-connecting as much data as possible with just $\nu$, $\dot{\nu}$
and $\ddot{\nu}$, such that there was no noticeable contamination by timing noise 
appearing in the resulting residuals. The resulting $n$ measurements are
shown in Figure~\ref{fig:indices}. Note that there is some indication that
$n$ is growing with time, at at rate of 
$\Delta{n}/\Delta{t} \simeq 0.02 \pm 0.01$\,yr$^{-1}$, resulting in a change
in $n$ of $\sim5$\% over 5.5\,yr. However, this effect is only
significant at the $2\sigma$ level. A similar analysis 
of 21 yr of data of the young pulsar
PSR B1509$-$58 showed long-term variations of $\sim$1.5\% in $n$, though
these were shown to fluctuate about an average value of $n$ \citep{lkgm05}.
Further timing observations of \psr\ will show if its value of $n$ is
changing secularly or is subject to timing noise variations as in PSR
B1509$-$58.

\section{\asca\ and \beppo\ Observations}
\label{sec:ascaobs}
In order to check the validity of our \rxte\ analysis, we 
re-analyzed the timing signal in two archival observations of the
pulsar obtained with the {\textit{Advanced Satellite for
Cosmology and Astrophysics}} (\asca) observatory \citep{tih94}. Details of
the first observation (1993 October) can be found in \citet{bh96c} and
\cite{gvb+00},
while details of the second observation (1999 March) are described in \citet{vgtg00}.
Using the prescription in the \asca\ Data Reduction 
Guide\footnote{http://heasarc.gsfc.nasa.gov/docs/asca/abc/abc.html}
and standard FITS tools, we extracted photons from the two gas-imaging
spectrometers (GISs) in the 3 - 8\,keV range from a $4'$ radius aperture 
surrounding the
source. We chose this energy range in order to preferentially select the
harder energy photons which mainly come from the pulsar rather than the softer
energy events having a high probability coming from the supernova remnant 
\citep{gvb+00}. We adjusted photon arrival times to the solar system barycenter and then 
performed a periodogram analysis on the
event data centered on the pulse frequency predicted by our \rxte\ ephemeris. We 
folded the data with 10 phase bins and detected a weak pulse in each observation.
We generated a $\chi^2$ periodogram in frequency space and identified the peak
$\nu$. To determine
the uncertainty in the measured frequencies, we performed a Monte Carlo simulation that created fake data sets
containing a periodic signal (at the $\nu$ and amplitude determined from the
periodogram) and Poisson noise. A periodogram was then performed on each noisy fake
data set and the best $\nu$ was recorded. We performed 500 iterations for each
observation and took the standard deviation as the $1\sigma$ uncertainty
on our $\nu$ measurement. Using this method, we determined that for the
1993 October observation, $\nu = 3.094516(2)$\,Hz and for the 1999 March 
observation, $\nu = 3.082859(3)$\,Hz.

To verify our \rxte\ results, we compared
the above \asca-measured frequencies, as well as two 1999 \beppo\ timing
observations reported on by \citet{mbb+02}, with the \rxte\ ephemeris prediction.
Both \beppo\ $\nu$ measurements agree with our ephemeris within
uncertainties, as do the two \asca\ frequencies. That the 1993 \asca\
measurement of $\nu$ agrees with our prediction is somewhat surprising,
since the ephemeris used to predict the value covers dates beginning seven years
later. This does not necessarily exclude the possibility of glitches having
occured in the interim however, since glitches in other young pulsars 
have been measured that sometimes have amplitudes $\sim 3$ orders of magnitude
smaller than we could detect given the uncertainties here \citep{lkg05}. 

If no glitches occurred during the 6-yr gap from 1993 to 1999, then it is
possible to fit all frequency measurements prior to the MJD 52210 glitch
with a second order polynomial in order to measure the braking index.
Performing this analysis, the resulting braking index is $n=2.44\pm 0.09$,
2.2$\sigma$ from our phase-coherent result. 

\section{Discussion}
\label{sec:discussion}
Establishing true ages of pulsars is important for
several aspects of neutron-star studies, including neutron star cooling, 
population synthesis, spatial velocity estimates, and to consider their 
associations with supernova remnants. Based on spin-down properties alone,
the true pulsar age is impossible to determine. However, we can estimate
the
age based on the standard spin-down model ($\dot{\nu} \propto -
\nu^n$) given our measurement of $n$. Integrating this model yields,
\begin{equation}
\tau= -\frac{1}{n-1} \frac{\nu}{\dot{\nu}} \left(1 -
\left(  \frac{\nu}{\nu_0}\right)^{n-1}   \right).
\end{equation}
Assuming that the initial spin frequency was much larger than the current value,
$\nu_0 \gg \nu$, the age estimate for \psr\ becomes,
\begin{equation}
\tau \le - \frac{1}{n-1} \frac{\nu}{\dot{\nu}} \la 884 \, \rm{yr}.
\label{eqn:age}
\end{equation}
Since the initial spin frequency is unknown, the above 
estimate provides an upper limit on the pulsar age. Therefore, assuming
that $n$ and $K$ have remained constant over the lifetime of the
pulsar, the upper limit on the age of \psr\ is $\sim$884\,yr. This value is
larger than the characteristic age of 723\,yr, but still smaller than the
unambiguously known age of the next youngest pulsar, that in the Crab Nebula.

\citet{mbb+02} did an analysis of frequency measurements of the \asca\ and
\beppo\ data, as well as a small subset of the \rxte\ data reported here. 
Although data gaps prevented them from a 
firm conclusion regarding $n$ because of the possibility of glitches,
their tentative estimate (made assuming no glitches nor substantial
timing noise) was $n=1.86 \pm 0.08$, inconsistent with our result. 
There are two plausible explanations for this discrepancy. The first is that
significant timing noise and/or glitches cannot be definitely excluded in
the 6-yr gap between the successive \asca\ observations. 
The second is that the coordinate used to barycenter these data was
ultimately found to be off by $\sim 7''$. The earlier study used coordinates
derived from \asca\ data with a $20''$ position uncertainty, prior to the
availability of the new measurement provided by the \textit{Chandra X-ray
Observatory} used herein. 

Our measured $n=2.65\pm0.01$ is significantly less than 3,
as is the case for all established values of $n$, which are shown
in Table~\ref{table:indices}. There are several ideas for 
the nature of the deviation from the prediction of simple magnetic dipole
braking. A time-varying magnetic moment could produce an observed
$n$ less than the true $n_0=3$ \citep{bla94}. A varying \textit{B}-field can 
in principle be verified
with a measurement of the third frequency derivative, $\nudotdotdot$, which
may or may not ultimately be possible for this source, depending on 
the number and strength of the glitches it experiences, as
well as on the strength of its timing noise. Presently, $\nudotdotdot$ has
been measured for only two pulsars, both of which are consistent with a
constant $B$ \citep{lps93,lkgm05}. Another suggested 
explanation for $n<3$ is that a fallback disk forms from supernova material
and modulates the spin-down of young pulsars via a propeller torque.
Spin-down via a combination of magnetic dipole radiation and propeller
torque results in $2<n<3$ \citep[e.g.][]{aay01}. However, this requires that
the disk material does not suppress the pulsed emission during
the propeller phase, which is difficult to achieve \citep{mph01}. 
Angular momentum loss due to a stellar wind would result in $n=1$ 
\citep{mt69}. Thus spin-down due to some
combination of relativistic pulsar winds associated with young
neutron stars and observed indirectly as pulsar wind nebulae
\citep[e.g.][]{rtk+03}, and magnetic dipole spin down may 
result in measured values $n<3$. 

\citet{mel97} presented an intriguing solution to the $n<3$ problem. 
He postulated that the radius pertinent to dipole
radiation is not the physical neutron-star radius, but a ``vacuum radius'',
associated with the location closest to the neutron star where field-aligned
flow breaks down. Since this radius can be significantly larger than the 
neutron-star radius, the system can no longer be treated as a point dipole
and $n=3$ is not necessarily true. 
His model provides a prediction for $n$, given three observables: $\nu$, $\dot{\nu}$
and $\alpha$, the angle between the spin and magnetic axis. This model predicts
that $2<n<3$ for all pulsars and that $n$ approaches 3 as a pulsar ages. 
Currently, there is no measurement of $\alpha$ for \psr. However, for the
model to be consistent with our measured $n$ within $3\sigma$, $\alpha$ must
lie between $8.1 \degrees$ and $9.6 \degrees$. 
This is small but not unreasonable given the very broad pulse profile for this source.
Future radio polarimetric observations could in principle constrain $\alpha$,
though at present, no radio detection of this source has been reported
\citep{kmj+96}. 

The measurement of $n$ for \psr\ brings the total number of measured braking
indices to six, shown in Table~\ref{table:indices}, along with $\nu$,
$\dot{\nu}$, $\tau$, $\tau_c$, $B_{\rm{dipole}}$ and $\dot{E}$ for each object. 
Comparing this new value to the other four measurements obtained via typical timing 
methods (i.e. excluding the Vela pulsar whose $n$ was measured using a
different method, due to the large glitches experienced by this pulsar), 
we find no correlations between $n$ and any of the other parameters.
If the Vela pulsar is included in the analysis, there is a slight 
correlation between $n$ and characteristic age. However, there is a relatively large
scatter among the five younger pulsars, suggesting the Vela pulsar's value
could also be just an extremum of the scatter.

The large scatter in the observed values of $n$ could be intrinsic to the
pulsars, for instance, there may exist a relationship between $n$ and
$\alpha$ or $B$, as suggested by \citet{mel97}. One way to test
this possibility would be to obtain more precise measurements
of $\alpha$ for the youngest pulsars, e.g. via polarimetric
observations. This, however, may prove difficult given that young radio 
pulsars tend to show very flat position angle swings 
\citep[e.g.][]{cmk01,jhv+05}. Alternatively, the scatter could be the result
of a physical process outside the neutron star that is affecting pulsar
spin down, such as a supernova fallback disk as suggested by \citet{mph01}.
A supernova fallback disk was recently discovered by \citet{wck06}
around a young neutron star, though there is no indication that the disk and
the neutron star are interacting. 

\psr\ is in an emerging class of high magnetic field rotation-powered
pulsars \citep[e.g.][]{ckl+00,pkc00,km05,msk+03}. The existence of these
sources raises the question of why they appear to be rotation powered instead of
powered by the decay of their large magnetic fields, as are the magnetars.
The magnetar model proposed by \citet{tlk02}
provides a prediction for the X-ray luminosity of a neutron star
with a magnetic field of $\sim 10^{14}$\,G given a measurement of $n$.
In the model, the neutron-star magnetosphere suffers a large scale `twist'
of the north magnetic hemisphere with respect to the southern hemisphere,
and resulting magnetospheric currents both scatter thermal surface photons
and themselves heat the surface upon impact. Both effects result in the
observed X-ray emission. The model predicts a one-to-one relationship
between twist angle and $n$, as well as $L_x$.
In the case of \psr, if the above model were applicable, the measurement of
$n=2.65+/-0.01$ would imply a twist angle of $\phi
\simeq 1.2$\,rad, which would predict an X-ray luminosity of
$L_x \simeq 3.7 \times 10^{35}$\,erg\,s$^{-1}$. 
The observed luminosity of \psr\ in the 0.5-10\,keV energy range 
is $L_x=4.1 \times 10^{35}$\,ergs\,s$^{-1}$ for a distance of 19\,kpc
\citep{hcg03}. This value is consistent with the magnetar model prediction.
This suggests that the pulsar may actually be a magnetar. However, 
the estimated distance is is uncertain by a factor of two and 
likely overestimated, since the remnant flux and size are much larger than
expected for the pulsar's age. Thus the true X-ray luminosity of \psr\ could
be significantly less than the magnetar model prediction, though this will
likely remain uncertain. In any event, since the observed X-ray flux can be
easily accounted for by the spin-down luminosity of the pulsar, \psr\ is
clearly not an `anomalous' X-ray pulsar. Moreover, the pulsar's X-ray
spectrum is much harder than that of any anomalous X-ray pulsar, and
inconsistent with what is predicted in the magnetar model. Indeed the
spectrum is much more in line with those seen from rotation-powered pulsars
\citep{got03}. 
More theoretical work needs to be
done to explain the difference between the magnetars and the rotation-powered
pulsars having inferred magnetar-strength fields.

\acknowledgments
This research made use of data obtained from the High Energy Astrophysics
Science Archive Research Center Online Service, provided by the NASA-Goddard
Space Flight Center. We thank F. Gavriil for useful discussions and an
anonymous referee for helpful comments. MAL is an NSERC PGS-D fellow. 
VMK is a Canada Research
Chair. Funding for this work was provided by NSERC Discovery Grant Rgpin 
228738-03.  Additional funding came from 
Fonds de Recherche de la Nature et des Technologies du Quebec, 
the Canadian Institute for Advanced Research, and the Canada Foundation
for Innovation. Funding has been provided by NASA \rxte\ grants over the 
course of this study. EVG acknowledges NASA ADP grant ADP04-0000-0069.

\clearpage

\normalsize

\normalsize

\normalsize

\normalsize

\begin{center}
\begin{deluxetable}{lc}
\tablecaption{Phase-coherent Timing Parameters for \psr. \label{table:ephem}}
\tablewidth{0pt}
\startdata
\hline
\multicolumn{2}{c}{First phase-coherent solution \tablenotemark{a}}
\\ \hline \hline
Dates (Modified Julian Day)        & 51574.2 -- 52837.4 \\
Dates (Years)                      & 2000 Jan 31 -- 2003 Jul 17 \\
Number of TOAs                     & 47 \\
Epoch (Modified Julian Day)        & 52064.0 \\
$\nu$ (Hz)                         & 3.0782148166(9) \\
$\dot{\nu}$ ($10^{-11}$~s$^{-2}$)  & $-$6.71563(1) \\
$\ddot{\nu}$ ($10^{-21}$~s$^{-3}$) & 3.87(2) \\
Braking Index, $n$                 & 2.64(1) \\
Number of derivatives fitted       & 8 \\
RMS residuals (ms)                 & 7.65 \\
Glitch epoch (Modified Julian Day)  & 52210(10)  \\
$\Delta{\nu}/{\nu}$                & $2.5(2) \times 10^{-9}$ \\
$\Delta{\dot{\nu}}/{\dot{\nu}}$    & $9.3(1)\times 10^{-4}$ \\
\hline
\multicolumn{2}{c}{Second phase-coherent solution \tablenotemark{a}}
\\ \hline \hline
Dates (Modified Julian Day)        & 52915.8 -- 53578.6 \\
Dates (Years)                      & 2003 Oct 3 -- 2005 Jul 27 \\
Number of TOAs                     & 31 \\
Epoch (Modified Julian Day)        & 53404.0 \\
$\nu$ (Hz)                         & 3.070458592(1) \\
$\dot{\nu}$ ($10^{-11}$~s$^{-2}$)  & $-$6.67793(5) \\
$\ddot{\nu}$ ($10^{-21}$~s$^{-3}$) & 3.89(4) \\
Braking Index, $n$                 & 2.68(3) \\
Number of derivatives fitted       & 5 \\
RMS residuals (ms)                 & 5.61 \\
\hline
\enddata
\tablenotetext{a}{Quoted uncertainties are formal $1 \sigma$ uncertainties
as reported by TEMPO.}
\end{deluxetable}
\end{center}

\begin{center}
\begin{deluxetable}{lccccccccc}
\tablewidth{0pt}
\tablecaption{Spin and inferred parameters for pulsars with measured
$n$ ordered by spin-down age\label{table:indices}}
\startdata
\hline
Name &  $n$\tablenotemark{a}  & $\nu$    & $\dot{\nu}$ & $\tau_c$ \tablenotemark{b} & $\tau$\tablenotemark{c}&  $B_{\rm{dipole}}$ \tablenotemark{d}& $\dot{E}$\tablenotemark{e} & Ref\tablenotemark{f} \\
     &      & (s$^{-1}$)& ($10^{-11}$s$^{-2}$) &  (yr) & (yr)& ($10^{12}$ G) & ($10^{36}$ erg s$^{-1}$) \\ \hline
     \hline
J1846$-$0258     & 2.65(1)  &3.07  & $-$6.68  &  723  &884 &  49  &   8.1 & (1)\\
B0531+21         & 2.51(1)  &30.2  & $-$38.6  &  1240 &1640&   3.8 & 460   & (2) \\
B1509$-$58     &2.839(3)   & 6.63  & $-$6.76  &  1550 &1690&   15 &  18   & (3)\\
J1119$-$6127    & 2.91(5)  & 2.45  & $-$2.42  &  1610 &1680&  42  &   2.3 & (4)\\
B0540$-$69      & 2.140(9) & 19.8  & $-$18.8 &   1670 &2940&   5.1 & 150   & (5) \\
B0833$-$45\tablenotemark{g}& 1.4(2)&11.2&$-$1.57&11300&57000&  3.4  &   6.9 & (6) \\
\hline
\enddata
\tablenotetext{a}{Uncertainties on $n$ are in the
last digit.}
\tablenotetext{b}{Characteristic age is given by $\tau_c = \nu/2\dot{\nu}$. }
\tablenotetext{c}{Inferred upper limit timing age given $n$,
$\tau = {\nu}/(n-1){\dot{\nu}}$, assuming $\nu_i >> \nu $. }
\tablenotetext{d}{Dipole magnetic field estimated by $B_{\rm{dipole}}=3.2 \times
10^{19}({-\dot{\nu}}/{\nu^3})$\,G, assuming $\alpha=90\degrees$ and $n=3$. }
\tablenotetext{e}{Spin-down luminosity,
$\dot{E} \equiv 4\pi^2I\nu\dot{\nu}$, where it is assumed that
$I=10^{45}$\,g\,cm$^2$
for all pulsars.}
\tablenotetext{f}{References (1) This work, (2) \citet{lps93}, 
(3) \citet{lkgm05}, (4) \citet{ckl+00}, (5) \citet{lkg05} (6) \citet{lpgc96}.}
\tablenotetext{g}{Braking index for the Vela pulsar was not determined from
a standard timing analysis due to the large glitches experienced by this object.
Instead, measurements of $\dot{\nu}$ were obtained from assumed `points of
stability' 100 days after each glitch \citep[see][, for details]{lpgc96}.}
\end{deluxetable}
\end{center}

\end{document}